\documentclass{article}
\usepackage{romp40}
\usepackage{amsmath}
\usepackage{amssymb}

\def \msk {\vskip 3mm}         
\def \bsk {\vskip 5mm}

\def \ph {{\varphi}}
\def \G {{\cal G}}
\def \H {\mbox{${\cal H}$}}

\def \L {{\cal L}}
\def \M {{\cal M}}
\def \O {{\cal O}}

\def \Z {{\cal Z}}

\def \LRM{{\Lambda_R(\M)}} 

\def\doppio#1{{\rm I}\kern-.1667em{\rm #1}}

\def\N{{\rm \doppio N}}

\def\R{{\rm \doppio R}}

\def\Q{\text{Q}\kern-.52em
    \text{\vrule height1.5ex width.5pt depth0pt}\kern.45em}

\def \cifm {C^\infty(\M)}

\def \difm { \mbox{Diff}(\M)}

\def \lrm {{\cal L}_R(\M)}

\def \GM {{\G(\M)}}

\def \glv {{g_{\lambda v}}}

\def \glzv {g_{\lambda z v}}

\def \Ulv {U(\lambda v)}

\def \Ulz {U(\lambda Z)}

\def \glv {g_{\lambda v}}

\def \gmw {g_{\mu w}}

\def\Z{{\mathchoice {\hbox{$\Ss\textstyle Z\kern-0.4em Z$}}
{\hbox{$\Ss\textstyle Z\kern-0.4em Z$}} {\hbox{$\Ss\scriptstyle Z\kern-0.25em
Z$}} {\hbox{$\Ss\scriptscriptstyle Z\kern-0.2em Z$}}}}

\def\C{{\mathchoice{\hbox{$\rm\textstyle\text{\kern.35em\vrule
   height1.5ex width.5pt depth0pt\kern-.35em C}$}}
{\hbox{$\rm\textstyle\text{\kern.35em\vrule
   height1.5ex width.5pt depth0pt\kern-.35em C}$}}
{\hbox{$\rm\scriptstyle\text{\kern.35em\vrule
   height1.5ex width.3pt depth0pt\kern-.35em C}$}}
{\hbox{$\rm\scriptscriptstyle\text{\kern.35em\vrule
   height1.5ex width.2pt depth0pt\kern-.35em C}$}}}}

\def \be {\begin{equation}}

\def \ee {\end{equation}}

\def \ra {\rightarrow}

\def \ume {{\scriptstyle{\frac{1}{2}}}}

\def \gotA {{\cal A}(\M)}

\title{ Classical and Quantum Mechanics from the universal 
Poisson-Rinehart algebra of a manifold }
\author{ G.Morchio \\ Dip. di Fisica, Universit\`a di Pisa and INFN, Pisa, Italy 
\\ e-mail: morchio@df.unipi.it \\[2ex]
          F. Strocchi 
                      \\ Scuola Normale Superiore, Pisa, Italy and INFN, Pisa
 \\ e-mail: f.strocchi@sns.it  }

\begin{document}

\maketitle

\begin{abstract}
The Lie and module (Rinehart) algebraic structure of vector fields
of compact support over $C^\infty$ functions on a 
(connected) manifold $\M$ define a unique universal non-commutative 
Poisson $*$ algebra $\Lambda_R(\M)$. 
For a compact manifold, a (antihermitian) variable $Z \in\LRM$, central
with respect to both the product and the Lie product,
relates commutators and Poisson brackets; in the non-compact case, 
sequences of locally central variables allow for the addition of
an element with the same r\^ole.
Quotients with respect to $ Z^* Z -  z^2 \,I$, $z \geq 0$, define 
classical Poisson algebras and quantum observable algebras, 
with $z = \hbar$. Under standard regularity conditions, the 
corresponding states and Hilbert space representations 
uniquely give rise to classical and quantum mechanics on $\M$.

\end{abstract}


\section{Introduction and results}

According to Heisenberg and Dirac, Quantum Mechanics is 
obtained from Classical Mechanics by replacing 
classical Poisson brackets, multiplied by the Planck constant,
by commutators.

Two circles of problems arise from such an \lq\lq ansatz\rq\rq :
one concerning the precise content of that prescription, which is not 
so well defined as it may appear, if it is to be interpreted as 
general and unique, and therefore independent from choices of coordinates, 
Hamiltonian, etc.; the second asking whether such a \lq\lq substitution\rq\rq\ has
to be interpreted as a mere prescription or rather follows from more fundamental 
principles.

Concerning the first kind of problems, it is well known that
the substitution procedure is completely well defined for a particle 
in euclidean space: when applied to cartesian coordinates and their 
conjugated momenta, it defines the Heisenberg algebra, which, assuming
exponentiability of the generators, gives rise to a unique 
$C^*$ algebra \cite{Slawny}, the Weyl algebra, 
with a unique Hilbert space representation continuous in the group 
parameters (von Neumann Theorem \cite{Thirring}). 

The situation changes completely if one considers 
classical mechanics on a manifold $\M$ and asks whether a similar, 
coordinate independent, construction provides a unique algebra, 
with a similar classification of representations.
In this case, geometrical structures play an essential r\^ole and different
strategies and constructions have been proposed.

\lq\lq Phase space\rq\rq\ quantization methods start from the classical phase 
space $T^*(\M)$  and try to associate an element $Q(f)$ of an operator algebra
to any classical variable, i.e. any (regular) function $f$ on $T^*(\M)$.
Requests which seem {\em a priori} reasonable are however found to be inconsistent:
$[Q(f), Q(g)] = i \hbar \{ f, g \}$, $[ \ , \ ]$ denoting the commutator
and $\{ \ , \ \}$ the classical Poisson brackets, is in fact incompatible
with $Q(g(f)) = g(Q(f))$, and also with linearity of $Q$, 
if irreducibility of the resulting algebra, or related conditions, are assumed
\cite{Ali}. 

Possible solutions are given either by restricting the correspondence $Q$ 
to suitable subsets of the classical variables (\lq\lq Geometric
quantization\rq\rq ), or by relaxing the relation 
between commutators and classical Poisson brackets, which is assumed to 
hold to order $\hbar$ (\lq\lq Deformation quantization\rq\rq ).
In both cases, the construction depends on the introduction of additional
structures, respectively geometrical and algebraic, and the result is not unique.

On the other side, \lq\lq Canonical quantization\rq\rq\ developed
into an analysis based only on the geometry of $\M$ and of its diffeomorphism 
group, and therefore into the study of the representations of the crossed product 
algebras $C^0 (\M) \times G$, $G$ a subgroup of Diff $(\M)$,
defined by the action of $G$ on $C^0 (\M)$. 

In the Segal approach \cite{Segal}, a maximality condition on $C^0(\M)$
reduces the analysis to the representations in $L^2(\M)$, with the Lebesgue 
measure.

In the approach of Mackey \cite{Mackey} and Landsman \cite{LandsmanCP}, 
$\M$ is assumed to be a homogenous space, $\M = G/H$, $G$ a
finite dimensional Lie group, $H$ a subgroup. 
Given $\M$, the resulting algebra and representations 
(classified by the representations of $H$) substantially 
depend on the choice of $G$; leaving aside the interest 
of the additional degrees of freedom associated to $H$, the construction 
does not therefore provide a unique formulation of Quantum Mechanics 
for a particle on $\M$.

Segal's diffeomorphism invariant approach has been developed 
by Doebner \cite{Doebner}, who dropped Segal's maximality condition
by assuming a local Hilbert space structure of any dimension, with a 
connection form which relates spaces at different points.

The representations of the diffeomorphism group, associated in 
general to diffeomorphism invariance, have been studied by 
Goldin.
The corresponding Lie algebra of functions and vector fields 
does not contain enough information for the identification 
of mechanics on $\M$; in fact, it has the
interpretation of the (classical) current algebra and its 
representations appear therefore in many situations, 
in particular for all $N$ particle quantum 
(Schroedinger) systems on $\M$ \cite{Goldin}.
The same considerations apply to the representations of 
the crossed product $C^*$ algebra  $C^0 (\M) \times $ Diff $(\M)$.

Clearly, the basic problem of the diffeomorphism invariant 
canonical approaches is the identification of degrees of freedom 
for the generalized momenta, which is not correctly given by the 
Lie structure of vector fields, since linearly independent 
vector fields define independent variables.

The solution proposed in \cite{MS1} is to consider as fundamental
the module structure of the Lie algebra of vector fields
of compact support, denoted
by Vect $(\M)$, on the algebra of $C^\infty$ 
functions, i.e. the {\em Lie Rinehart} (LR) structure of 
($C^\infty (\M),$ Vect $(\M)$) and to assume that the Lie-Rinehart 
product 
\be
f \circ v: (C^\infty(\M) , {\rm Vect}(\M)) \rightarrow {\rm Vect}(\M)
\ee
is realized, in the algebra defining Quantum 
Mechanics (QM) on $\M$, by the {\em symmetric ( Jordan) product\/}: 
\be 
f \circ v = 1/2 \, (f \cdot v + v \cdot f)
\label{LR}
\ee

It turns out that the LR relations (\ref{LR}) can also be written 
in terms of the resolvents of the unbounded operators 
representing vector fields (of compact support) and define 
therefore, together with the crossed product relations 
between $C^\infty (\M)$ and Diff $(\M)$, a unique $C^*$ algebra. 
Its Hilbert space representations have been classified, assuming regularity,
i.e strong continuity of one dimensional subgroup of Diff $(\M)$ as for the
Weyl algebra, and shown to be in one to one correspondence with
the unitary representation of the fundamental group of $\M$, describing
the displacement of a particle along non-contractible closed 
paths \cite{MS1}. 
Such a classification of states reproduces, only assuming basic 
geometrical and algebraic structures, that obtained by Doebner
\cite{Doebner} and by \cite{Zanghi}, the latter within an approach 
{\em a priori} based on trajectories. 

For the basic questions about the nature of quantization,
clearly one has to identify general principles and ask to which extent they
constrain both classical and quantum mechanics and which alternatives they 
leave open.

The strategy proposed by Dirac \cite{Dirac}, with his analysis of 
\lq\lq proportionality between commutators and Poisson brackets\rq\rq ,
ends with the difficulties of phase space quantization.
As we shall see, Dirac's equations cannot be interpreted as directly 
relating classical and quantum algebras and the basic missing point 
is the very identification of the algebraic structures to which they apply.

We start therefore from fundamental principles, given by the geometry 
of $\M$ and Vect $(\M)$. 
They are embodied in the commutative algebraic structure of 
$C^\infty (\M)$, describing the manifold, and in the Lie structure of
$C^\infty (\M) + $ Vect $(\M)$ defined by the Lie relation between vector fields
and their action on $C^\infty (\M)$. From the above discussion it is clear
that also the module structure of Vect $(\M)$ over $C^\infty (\M)$ plays
an essential r\^ole, redefining linear dependence of vector fields according
to multiplication by $C^\infty$ functions.

Actually, the Lie-Rinehart algebra ($C^\infty (\M),$ Vect $(\M)$) 
is represented {\em faithfully} both in classical and 
in quantum mechanics, with the Lie product realized 
respectively as the Poisson and commutator brackets and the LR product 
realized as the symmetric product, eq.(\ref{LR}).

We then propose to base the most general notion of mechanics 
on a set of variables indexed by $C^\infty (\M)$  and  Vect $(\M)$) 
with their Lie-Rinehart relations, $C^\infty$ functions 
being interpreted as position variables and  Vect $(M)$, 
describing \lq\lq small displacements\rq\rq , as 
\lq\lq generalized momenta\rq\rq . 

In order to obtain variables to which well defined values may be 
assigned in terms of a notion of spectrum, 
we consider (associative) algebras generated by them. 
The associative product is assumed 1) to extend the commutative 
product of $C^\infty(\M)$ and 2) to reproduce, with its 
{\em symmetric (Jordan) part}, the Lie-Rinehart product between  
$C^\infty (\M)$ and Vect $(\M)$, i.e. to satisfy eq.(\ref{LR});
as discussed above in the case of QM, condition 2) has a basic r\^ole
for the identification of degrees of freedom and therefore for the 
characterization of Mechanics on $\M$, with respect to the most
general diffeomorphism invariant system. The use of  {\em the symmetric part}
of the associative product is essential in the non-commutative case.

We also assume 3) that the Lie product on 
$C^\infty (\M) + $ Vect $(\M)$ can be extended to a Lie product on 
such algebras, defining on them derivations, i.e., satisfying the Leibniz rule 
with respect to the associative product. This may be interpreted as the
association of some (infinitesimal) operation to each variable, 
generalizing the action of vector fields and allowing for a general notion
of symmetry transformation (which is essential, e.g., for the
introduction of a time evolution).

We extend therefore the Lie-Rinehart algebra of $\M$
to a {\em non-commutative (real) Poisson algebra}. In general, one obtains an
enveloping non commutative Poisson algebra of a Lie-Rinehart algebra \cite{MS2},
a notion which extends that of Poisson
enveloping algebra of a Lie algebra \cite{Voronov}.
It should be emphasized that no relation is assumed between Lie 
products and commutators, only the Leibniz rule constraining
the associative and Lie products.

If no other restriction is assumed, the result is the
{\em universal} enveloping (non-com\-mu\-tative) Poisson algebra of the
LR algebra $(C^\infty (\M)$, Vect $(\M))$. Its uniqueness follows from the 
definition of universality (see below) and its construction 
has been given in \cite{MS2}. 

Such a non-commutative Poisson algebra will be called the 
Lie-Rinehart universal Poisson algebra of $\M$, or briefly the 
{\em Poisson-Rinehart} algebra of $\M$ and denoted by $\Lambda_R(\M)$. 
A unique linear involution is also defined (see Section 2) on 
$\LRM$, treating functions and vector fields as real variables, 
$f = f^*$, $ v = v^* $ and satisfying $(A B)^* = B^*\,A^* $, 
which then implies $\{\,A,\,B\}^* = \{\,A^*,\,B^*\,\}$.

{\em A priori}, $\LRM$ is a very general algebraic structure, 
in fact the most general associative algebra, with a 
Lie bracket satisfying the Leibniz rule, generated by the LR 
algebra associated to $\M$ and Vect $(\M)$. The main result is that it 
describes {\em nothing else than classical and quantum mechanics}. 

This is obtained as follows: first, we recall that in general, 
in a Poisson algebra $\Lambda$, commutators and Lie products 
satisfy the following relation, already pointed out by Dirac 
\cite{Dirac} and rederived in refs. \cite{Voronov}~\cite{Farkas}: 
\be
{[\,A,\,B\,]\,\{\,C, \,D\,\} =
\{\,A,\,B\,\}\,[\,C,\,D\,] \ \ \ \ \forall A, B, C, D \in \Lambda \, .}
\label{Farkas}
\ee
Clearly, if for some $C,D \in \Lambda$, $\{\,C, \,D\,\}$ has an inverse, eq.
(\ref{Farkas}) allows to express commutators in terms of Poisson brackets.
More generally, the same holds for prime Poisson algebras, i.e. 
algebras without ideals which are divisors of zero~\cite{Farkas}; such 
conditions are not satisfied by $\LRM$, since (see below) any pair of 
functions with disjoint supports generate (bilateral) ideals $I_1, I_2$ with 
$I_1\cdot I_2 = 0$. 
However, in the case of compact $\M$, by summing commutators of locally 
conjugated variables in $\LRM$, we construct  a unique variable 
$Z \in \LRM$ satisfying
\be
   [A, B] = Z \{A, B\}  \ \ \ \  \forall A, B \in \LRM  \,  .
\label{Z}
\ee
$Z$ turns out to be central with respect to both the associative 
and to Lie product; in the non compact case, we construct 
a sequence $Z_n$ satisfying eq.(\ref{Z}) for $n \geq \bar n (A,B)$, 
allowing for an extension of $\LRM$ by an element satisfying eq.(\ref{Z}) 
and central in the same sense. In both cases, $Z$ is antihermitian:
$Z^* = -Z$.

If the variable $Z$ is substituted by an (imaginary) number, the 
result may be seen as a precise version of an argument by Dirac, based on 
eq.(\ref{Farkas}), on the commutator prescription for QM; 
more properly, it shows that the general approach to mechanics on a 
manifold provided by the basic geometrical (LR) structure automatically 
yields a variable invariant under diffeomorphisms and under all 
the physical operations, expressed  by the Lie brackets in $\LRM$,
i.e. a \lq\lq universal constant\rq\rq,  with the same r\^ole and 
interpretation as the Planck constant.

To proceed with the analysis of $\LRM$, one considers
the ideals generated by  $Z^* Z - z^2 I$, $z \geq 0$,
$I$ the identity of $\LRM$, given by $1 \in C^\infty(\M)$; 
they are stable with respect to the Poisson brackets 
with $\LRM$ and define therefore homomorphisms 
$\pi_z$ and quotient Poisson algebras $\LRM_z \equiv \pi_z (\LRM)$.

For $z = 0$, one obtains the commutative Poisson algebra generated
by $\cifm$ and by the $C^\infty$ vector fields, with the
natural module structure of Vect $(\M)$ on $\cifm$, which is isomorphic 
to the commutative Poisson algebra of polynomials in the 
cotangent vectors on $\M$ with $C^\infty$ coefficients;
under standard regularity conditions (Section 2), 
it has a unique Hilbert space representation, 
by multiplication operators in $L^2 (T^*(\M))$,
with Lie brackets represented by the classical Poisson 
brackets. $T^*(\M)$ arises as the spectrum, modulo a zero 
measure subset, of the commutative $C^*$ algebra 
generated by $C^\infty$ functions and exponential of vector fields.

For $z > 0$, $\pi_z(Z) = \iota z$, $\iota^2 = -1$ and there is
an isomorphism $\ph$,  mapping the real Poisson involutive algebra 
$\LRM_z $ into the
complex algebra generated by $C^\infty(\M)$ and by the generalized momenta
$T_v$ associated to the vector fields of Vect ($\M$), 
satisfying $$[\,T_v, \,T_w
\,] = i\, z\, T_{\{ v, \,w \}}\, , \ \ \ [T_v, \,f ] = i\,
 z\, \{ v , \, f\} \, , 
\ \ \ T_{f \circ v} = 1/2 \, (f T_v + T_v f) \, ,$$ with
 $\ph(f) = f$, $\ph(v) = T_v$, $\ph(\iota) = i.$
This is the (unbounded,  \lq\lq Lie-Rinehart\rq\rq ) 
quantum algebra introduced in ref. \cite{MS1}. 
Its regular (i.e. exponentiable) Hilbert space representations 
were studied in \cite{MS1} and shown to be in one-to one 
correspondence with the unitary representations of the fundamental group of
$\M$, $\pi_1(\M)$.

The analysis of Hilbert space representations 
of $\LRM$ shows that the above classification 
is complete, i.e. Classical and Quantum Mechanics on $\M$ are 
the only {\em regular} and {\em factorial} representations of $\LRM$ 
(Section 2).

The isomorphism between the real algebra $\LRM_z$, $z \neq 0$, 
and the above complex algebra also explains the origin of a complex 
structure in the standard formulation of quantum mechanics, through
a non-zero complex number representation of the antihermitian variable 
$Z$; no complex structure arises in classical mechanics, which originates 
from the zero representation of $Z$.

For $\M = {\bf R}^n$, if
only cartesian vector fields are considered, a similar simplified 
construction applies, still providing a $Z$ variable and the above
classification. 
The relevance of the full Rinehart structure is in fact only related to the 
geometrical identification of observables in a diffeomorphism 
invariant formulation (a question in fact at the origin the above 
mentioned difficulties and alternatives for the formulation of QM
an manifolds).

In all cases, the Lie, or Lie-Rinehart, algebra of momenta and 
(functions of) positions is {\em the same} in classical and in 
quantum mechanics. The classical-quantum alternative only arises 
when {\em polynomials in the momenta} are introduced
and is uniquely given by the values of $Z$ in the universal enveloping
algebra, with the LR constraint on the symmetric product.

The above analysis unravels the basic r\^ole of 
the LR geometry of the configuration manifold, which in the quantum 
case is somewhat hidden in the observable $C^*$-algebra (Lie products 
being identified with commutators) and in the classical case 
goes beyond the abelian algebraic relations.
The LR algebraic structure provides, through the Poisson-Rinehart 
algebra $\LRM$, a notion of {\em non commutative phase space\/} which 
coincides with that of a {\em general mechanical system\/}, exactly covering 
classical and quantum mechanics.

In particular, the above construction shows that the Dirac ansatz 
of canonical quantization, in the form of the proportionality of the commutators 
{\em of variables in\/} $C^\infty (\M) +$ Vect $(\M)$ to their 
classical Poisson brackets, has no alternative, within 
the above rather general notion of mechanical system. 

The uniqueness of the commutators for $C^\infty$ functions and 
vector fields on the configuration manifold also explains the 
obstructions which arise by requiring proportionality 
of commutators to Poisson brackets {\em for all functions\/} 
on the classical phase space.
In our approach, the extension of the commutation relations 
starting from $C^\infty(\M)$ and Vect($\M$) and the construction of 
quantum algebras does not in fact use the classical Poisson algebra; it 
is given by the Leibniz rule and by the identification of the LR 
product with the symmetric product and therefore, in a sense, 
it {\em automatically depends\/} on $Z$.

Our results suggest a quite different approach 
to the relation between Classical and Quantum Mechanics with respect to phase space 
quantization: the classical phase space is {\em not} assumed as a starting point 
and rather arises from the same (non-commutative) Poisson algebra
in correspondence with one of the values taken by the central variable $Z$,
on the same footing as the quantum mechanical state space.

We also emphasize that in the above approach the Planck constant 
{\em needs not to be introduced}. It automatically appears as 
a variable invariant under all physical transformations, 
i.e. a {\em universal constant}, in the Poisson-Rinehart algebra 
of a manifold.

In the following Section the above notions will be formalized,
together with their implications 
on classical and quantum mechanics.

\section{The Poisson-Rinehart algebra of a manifold and its representations}

\msk\noindent
{\em The Lie-Rinehart algebra of $\M$}. 

\msk
A general notion of mechanical variables on $\M$, should include
regular ($C^\infty$) functions and vector fields, indexing generalized 
momenta; as we shall see, local variables are enough for a
general notion of Mechanics on $\M$.
We therefore consider the algebra generated by real functions 
of compact support in $\M$ and the identity, $C^\infty(\M)$,  and  
the space Vect ($\M$) of $C^\infty$ vector fields of compact support. 
Vect ($\M$) is a Lie algebra of derivations $v: f \rightarrow v(f)$ 
on $C^\infty(\M)$,
with Lie product $\{ \, v , \, w \, \}$ defined by
\be
\{ \, v , \, w \, \} (f) = v(w(f)) - w(v(f)) \ ;
\ee
its elements are integrable to one-parameter subgroups of the 
diffeomorphism group Diff ($\M$) by compactness of their support and
generate a subgroup of it, $\GM$.
As a real vector space, Vect ($\M$) is generated by an infinite number 
of linearly independent vector fields, which define independent 
variables; however, Vect ($\M$) is also a module over 
$C^\infty (\M)$ and, as such, it is locally 
generated by $n$ vector fields, $n$ the dimension of $\M$. 
The module structure of vector fields over $C^\infty(\M)$  
is clearly an expression of the functional character of the Lie
algebra Vect ($\M$), to which a notion of linear dependence with 
functions as coefficients is naturally associated; 
clearly, it is crucial in order to describe the infinite dimensional 
diffeomorphism group and its Lie algebra in terms of a finite 
number of generators, which will be interpreted
as independent generalized momenta.
 
All together, the above algebraic structures give rise to  
the {\em Lie-Rinehart algebra of} $\M$, $\lrm $, defined \cite{Rinehart} as
the pair  ($C^\infty (\M), $ Vect $(\M)$) with the commutative (real)
algebraic structure of $C^\infty(\M)$, the Lie product in Vect ($\M$), the
action of Vect ($\M$) on $C^\infty(\M)$ as derivations and the Rinehart
product $ (C^\infty(\M) , {\rm Vect}(\M)) \rightarrow {\rm Vect}(\M)$,
defined by its action as a derivation on  $C^\infty(\M)$:
\be
(f \circ v) (g) = f \, v(g) \ .
\ee
The Rinehart product is distributive in both factors, associative in the first,
\be
f \circ (g \circ v) = (f g) \circ v 
\ee
and is related to the Lie product by
\be
\{v, \,f \circ w\} = v(f) \circ w + f \circ \{v, \,w\} 
\ee 
for all $f,g \in C^\infty(\M)$, $v,w \in $ Vect ($\M$).   
The identity $1$ of $C^\infty(\M)$  satisfies 
$ v (1)  = 0 $, $1 \circ v = v$, $\forall v \in$ Vect ($\M$).

The action of Vect ($\M$) on $C^\infty(\M)$ as derivations 
can also be written as an extension of the Lie product Vect $(\M)$
to $C^\infty (\M) + $ Vect $(\M)$, which becomes therefore a Lie algebra,
still denoted by $\lrm$.
\be
\{ v , f \} \equiv v(f) \ , \ \  \{ f, g \} \equiv 0 
\ee
for all $f,g \in C^\infty(\M)$, $v \in $ Vect ($\M$).   

$\difm$ defines a group of automorphisms of the Lie-Rinehart algebra
$\lrm $. The action of the one-parameter group $\glv$, $\lambda \in \R$, 
generated by $v \in$ Vect ($\M$) satisfies
\be 
 (d/d\lambda)\, \glv (A)
= \{\,v , \,\glv(A) \, \}, \ \ \ \forall A \in C^\infty (\M) + {\rm Vect} (\M)
\label{ActDiff}
\ee
with the derivative taken in the $C^\infty$ topology.

\bsk\goodbreak\noindent
{\em Non-commutative Poisson $*$ algebras}.

\msk
As discussed in the Introduction, in order to obtain variables 
taking well defined values through a notion of spectrum, 
multiplications  should be allowed and an associative
algebra $\Lambda$ should be considered. 
In order to preserve diffeomorphism invariance, see eq.(\ref{ActDiff}), 
the Lie action of vector fields on $C^\infty (\M) + $ Vect $(\M)$,
should extend to derivations of $\Lambda$. 

If the interpretation of vector fields as generators of a symmetry
can be extended to all the variables in $\Lambda$, one is led to assume 
that their action is described by an extension to $\Lambda$ of the 
Lie product of $\lrm$ satisfying the Leibniz rule, 
in both arguments as a consequence of antisymmetry. 
Substantially, this is the step advocated by 
Dirac by the introduction of {\em generalized Poisson brackets},
assumed \cite{Dirac} to satisfy the Leibniz rule in an associative algebra.
Moreover, a notion of reality should be defined in $\Lambda$
through an involution, leaving $\lrm$ pointwise invariant. 

$\Lambda$ should therefore have the structure of a 
{\em non-commutative Poisson $*$ algebra} .
Non-commutative Poisson algebras have been formally introduced in refs. 
\cite{Voronov} \cite{FGV} \cite{Farkas} \cite{Dubois}.
They are real associative algebras, with product denoted by 
$A \cdot B$, which are also Lie algebras, with Lie product, 
denoted by $\{A,\,B\}$, satisfying  the Leibniz rule 
\be
{\{A, \,B \cdot C\}= \{A, \,B\} \cdot C + B \cdot \{A,\,C\},}
\ee
For a $*$ algebra, a linear involution must be defined, 
satisfying, as usual, $(A \cdot B)^* = B^* \cdot A^*$;
the reality of the Lie structure in $\Lambda$ also requires
$\{\,A , \,B\}^* = \{\,A^* , \, B^*\,\}$.

No relation is assumed between the Lie product and the 
commutator $[\,A,\,B\,] \equiv A \cdot B - B \cdot A$; 
however, the following identity holds for all $A,B,C,D$ in a 
Poisson algebra: 
\cite{Dirac} \cite{Voronov} \cite{Farkas}: 
\be
[\,A,\,B\,]\,\{\,C, \,D\,\} =
 \{\,A,\,B\,\}\,[\,C,\,D\,] \ .
\label{DVF}
\ee

\bsk\noindent
{\em The universal Poisson-Rinehart algebra of $\M$}.

\msk
Following the above arguments, a general notion of mechanics 
on a manifold $\M$ is given by the Poisson $*$ algebra 
{\em generated by the Lie-Rinehart algebra of} $\M$. 
More precisely, we consider the (non-commutative) 
{\em universal enveloping Poisson algebra} of the LR algebra 
$\lrm $, defined as follows \cite{MS2}:

\begin{definition}{Definition} 
The {\bf LR universal Poisson algebra}, or 
{\bf Poisson-Rinehart algebra}, of a manifold $\M$ 
is the unique (non-commutative) Poisson algebra $\LRM$  
with an injection  $i: \lrm \ra \LRM$ satisfying, 
\newline i) $i$ is a Lie algebra homomorphism,  
\be
i( \{l_1, \,l_2\}) = \{ i(l_1),
\,i(l_2)\} \, , \ \ \  \forall l_1, \,l_2 \in  \lrm  \ ,
\ee 
\newline ii)  
$i(1) \cdot i(l) = i(l)$, $\forall\, l \in \lrm)$,
\newline iii)  $i(f g) = i(f) \cdot i(g) \ \ \ \forall\, f, g \in \cifm$,
\newline iv) $i(f \circ v) = \ume (i(f) \cdot i(v) + i(v) \cdot i(f)) \ \ \
\forall\, f \in \cifm , \, v \in {\rm Vect} (\M)$
\newline
and such that, if $\Lambda$ is a Poisson algebra with injection $i_\Lambda$ 
satisfying i) - iv), there is a unique homomorphisms of 
Poisson algebras $\rho : \LRM \ra \Lambda  $ 
intertwining between the injections, $i_\Lambda = \rho\, \circ\, i \; $.
\label{LRM}
\end{definition}

As in general for enveloping algebras, the uniqueness of the  
Poisson universal enveloping algebra of $\lrm$ follows 
immediately from the uniqueness of the homomorphism $\rho$. 
In the following, $\lrm$ will be identified with the image 
of its injection in $\LRM$.

In order to construct $\LRM$, one may start from the 
Poisson universal enveloping algebra of $\lrm$ {\em as a Lie algebra}, 
introduced in general by Voronov \cite{Voronov} for (graded) Lie algebras,
and take quotients with respect to the ideals (in the sense of associative
algebras) generated by the relations ii)-iv); such ideals are in fact  
stable under the bracket operations with all the elements 
in the universal enveloping algebra of $\lrm$ as a Lie algebra, 
and therefore the corresponding quotients define Poisson algebras. 

The only delicate point in the construction of $\LRM$ 
is the validity of the Leibniz rule {\em in both arguments} 
for the extended the Lie brackets.
In fact, the Leibniz rule on one side determines a unique
extension of the Lie brackets to the tensor algebra of a Lie algebra.
The Leibniz rule on the other side is obtained by Voronov through 
an explicit analysis of the quotient with the ideal generated
by eq. (\ref{DVF}). The same result also follows
by imposing antisymmetry, through the quotient with respect to the 
ideal generated by all the elements $\{\,A,\,B\,\} + \{\,B,\,A\,\}$ 
together with their repeated $\{\,\cdot, \,\cdot\,\}$ brackets 
with the tensor algebra; the Jacoby identity follows by induction. 
Injectivity of $i$ holds since the above ideals have $0$ 
intersection with $i(\lrm)$.

With respect to the Poisson universal enveloping 
algebra of $\lrm$ as a Lie algebra, $\LRM$ includes 
the relations ii) - iv), so that, according
to the requirements discussed above, it extends the algebraic 
relations of $\cifm$ and the Lie-Rinehart product, identified with 
the symmetric (Jordan) part of the associative product.

Such relations are {\em a priori} essential for the mechanical interpretation of 
$\LRM$ and will in fact be crucial for the derivation of the classical phase 
space and for the characterization of QM on $\M$. 
They also enter in the construction of the Planck constant as a central
variable, even if conditions i) and ii) are sufficient for compact $\M$.

If, in Definition 1,  only conditions i) is assumed, the result
merely embodies the Lie relations between vector fields and functions
on $\M$, so that it applies in general to $\difm$ invariant systems; in
particular, the resulting Poisson algebra appears in all $N$-particle 
systems on $\M$~\cite{Goldin}, with $\lrm$, {\em as a Lie algebra}, 
interpreted as the current algebra.

On $\LRM$ there is a unique involution which leaves $\lrm$ pointwise 
invariant; in fact,  $(A \cdot B)^* = B^* \cdot A^*$ uniquely 
extends the involution from $\lrm$ to its 
tensor algebra, where it leaves invariant the ideals defining $\LRM$;
the involution is therefore well defined in $\LRM$ and, 
by construction of the Lie brackets in $\LRM$, satisfies
$\{\,A , \,B\}^* = \{\,A^* , \, B^*\,\}$.

$\difm$ is a symmetry of all the above constructions and 
extends therefore to a group of automorphisms of $\LRM$ as a 
Poisson algebra (leaving $\lrm$ invariant).
As before, the action of the one-parameter groups 
$\glv$, $\lambda \in \R$, satisfies eq.(\ref{ActDiff}),
for all $ A \in \LRM $, 
with the derivative taken in the topology induced on $\LRM$ by the 
$C^\infty$ topology on the tensor algebra over $\lrm$.
 
%\be 
% (d/d\lambda)\, \glv (A)
%= \{\,v , \,\glv(A) \, \}, \ \ \ \forall A \in \LRM) \, , v \in {\rm Vect} (\M)
%\label{ActDiffL}
%\ee
 
It should be noted that $\LRM$ is not an enveloping algebra in the usual sense 
\cite{Dix}, since the Lie product is not given by the commutator.
With respect to the commutative and non commutative 
Poisson algebras discussed in the literature 
\cite{FGV} \cite{Dubois} for classical mechanics and for quantum mechanics,
the concept of (non-commutative) Poisson universal enveloping algebra 
is more general and only includes the basic geometrical 
(Lie and Lie-Rinehart) structures. Its construction 
{\em includes neither classical nor quantum principles}, 
which are usually assumed in the form of abelianess 
or commutation relations.

\bsk\noindent
{\em The relation between commutators and Lie products}.

\msk\noindent
A central result in our analysis is the construction of a variable
$Z$ which relates commutators and Lie products in $\LRM$.
The essential ingredient is that any function of compact support, 
in particular the identity for compact $\M$, can be obtained as
a sum of Lie products; by eq. (\ref{Farkas}), the corresponding 
sum of commutators gives the required variable, which is then 
shown to be independent of the construction and central, both in the 
commutator and in the Lie sense. More precisely we have

\begin{theorem}{Theorem}  \label{TH1} \cite{MS1}
For a compact manifold $\M$, there exists a unique $Z \in \LRM$,
such that, $\forall A, \,B \in \LRM$, 
\be
[\, A, \, B \,] = Z\,\cdot \{\, A,\, B\, \} \ .
\label{Z1} \ee 
It satisfies
\be
\{\, Z,\, A\,\} = 0 = [\,Z,\,A\,] \ , \  \ Z = -Z^{*}
\label{Z2} \ee
For a non-compact manifold, there exists a sequence $Z_n = - Z_n^{*}\in \LRM $ 
such that, $ \forall A, \,B \in \LRM$, $\exists\, \bar{n}(A, B) \in \N $, 
such that, 
\be
[\, A, \, B \,] = Z_n\,\cdot \{\,A,\,B\,\} \  , \ \  \{\, Z_n,\, A\,\} = 0 =
[\,Z_n,\,A\,] \ \ \  \forall n > \bar{n} \ .  
\ee
One may therefore define an element $Z = - Z^{*}$, such that the Poisson
algebra $\tilde\Lambda_R (\M)$  generated by $\LRM$ and $Z$ satisfies eqs.
(\ref{Z1}), (\ref{Z2}).
\end{theorem}\goodbreak

\noindent
{\em Proof}. 
The proof simplifies for compact $\M$. In this case, 
the manifold can be covered by a finite number of open sets $\O_i$ 
homeomorphic to discs. There are therefore functions $q_i$ and vector 
fields $w_i$, with compact support contained (in local coordinates) 
in larger discs $\O'_i$, satisfying
$\{ q_i, w_i \} (x) = 1 \, , \ \forall x \in \O_i$.
For any partition of the unity $ \sum_i g_i = 1 $,  with Supp $(g_i) 
 \subset \O_i$, we have
\be
 1 = \sum_i \, g_i \{ q_i, w_i \}  = 
 \sum_i \, \{ q_i, g_i \circ w_i \} \equiv  \sum_i \, \{ q_i, p_i \} 
\ee
Then, eq.(\ref{Farkas}) gives, for all $A, B \in \LRM$,
\be
[\,A,\,B\,] = 1 \cdot [\,A,\,B\,]  
=  \sum_i \, \{q_i, p_i\} \cdot [ A,\,B ] =
\sum_i \, [\, q_i, \, p_i \, ] \cdot \{A,\,B\}  
\label{sqp}
\ee
The sum in the r.h.s. of eq.(\ref{sqp}) is independent of the
construction since, for any other choice of $\tilde \O_i , \tilde g_i, 
\tilde q_i, \tilde w_i$, eq.(\ref{sqp}) gives
\be
\sum_i \,  [\, \tilde q_i, \, \tilde p_i \, ] = 
\sum_j \,  [\, q_j, \, p_j \, ]  \cdot \sum_i \{ \tilde q_i, \, \tilde p_i \} =
\sum_j \,  [\, q_j, \, p_j \, ] \ .
\ee
One may therefore define 
\be
Z \equiv \sum_i  [\, q_i, \, p_i \, ] 
\ee
and eq.(\ref{Z1}) holds. By definition of the involution in $\LRM$, 
$Z = - Z^{*}$. By the Leibniz rule, $\forall A \in \LRM$,
\be
\{ Z, \, A \} =  
\sum \{ [ q_i, \,p_i ], \,A\}  =   
\sum ([ \{q_i, \,A\}, p_i ]  + [ q_i, \{\,p_i, \,A\} ]) \ ;
\label{ZA}
\ee
using eq.(\ref{Z1}) and the Jacobi identity for the Lie
product, the r.h.s. of eq.(\ref{ZA}) becomes
\be
Z \cdot \sum (\{\,\{q_i, \,A\}, p_i\} + \{\,q_i,\,\{\,p_i, \,A\,\}\,\} \,) 
= Z \cdot \sum \{\,\{q_i, \,p_i\}, \,A\} = Z \cdot \{ 1, \, A\} = 0 \ .
\ee
Then, eq.(\ref{Z1}) implies $[ Z, \, A ] = 0$, 
$\forall A \in \LRM$. $\rule{5pt}{5pt} $

Theorem 1 can be regarded as an answer to the problem raised by Dirac
 \cite{Dirac}  about the origin and the uniqueness of the relation 
between commutators in Quantum Mechanics and classical Poisson brackets. Dirac
 introduced the notion of {\em generalized Poisson brackets} 
(substantially, the notion of non-commutative Poisson algebra) 
as the basis for a generalization of Classical Mechanics and argued that 
Poisson brackets must be proportional to commutators on the 
basis of eq.(\ref{Farkas}); however, the argument relies on the 
\lq\lq independence\rq\rq of $C,D$ from $A,B$ in eq.(\ref{Farkas}) 
and is not conclusive, since invertibility of $\{ C, D\}$ is not discussed, 
and in fact the conclusion, even in the generalized form given
by eq.(\ref{Z1}), does not hold in general in (non-commutative) Poisson algebras.
E.g., one can derive, for the universal enveloping Poisson 
algebra of a finite dimensional Lie algebra $\L$, the relation
\be
[A,B] \cdot Z_1 = \{ A , B\} \cdot Z_2
\ee
\be
Z_1 = \sum_{ij} g_{ij} L_i \cdot L_j \ , \ \ \ 
Z_2 = \sum_{ijk} c_{ijk} L_i \cdot L_j \cdot L_k \ ,
\ee
with $g_{ij}$ the Killing form and $c_{ijk}$ the structure constant of $\L$;
$Z_1$ and $Z_2$ are central in the sense of Theorem 1, but
$Z_1$ is not invertible and, in general, $A \cdot Z_1 = 0$ 
does not imply $A=0$. 
For the Poisson-Rinehart algebra of a manifold, eq.(\ref{Z1})
holds as a consequence of condition ii) in Definition 1 for $\M$ compact and
from conditions ii) - iv) in general.

Moreover, Dirac's analysis is not conclusive because it is
unclear  {\em to which algebra} it is meant to apply,
so that the r.h.s. of eq.(\ref{Z1}) is {\em not\/} well defined.
If it is identified, as perhaps implicit in Dirac's analysis, with the
classical bracket in the {\em classical} Poisson algebra 
as a Lie algebra, leaving undetermined the associative product, 
one exactly meets the problems of {\em phase space quantization}.
If it is interpreted in the universal Poisson enveloping algebra of 
the {\em Lie algebra} of functions and vector fields, substantial 
information is lacking for the derivation of Quantum (and Classical) 
Mechanics on $\M$, as discussed above, and a conclusion can be obtained only
in fixed coordinates, e.g. in $\R^n$ with cartesian coordinates (see below).

The Lie-Rinehart relations and the construction of the 
Poisson-Rinehart universal enveloping algebra,
with the LR product identified with the Jordan product, is 
therefore essential for the relevance of eq.(\ref{Z1}); in particular, the
problems of phase space quantization are avoided since 
the construction of the classical Poisson algebra, also as a Lie algebra,
depends on abelianess of the product, which does not hold in $\LRM$;
in fact, the Lie algebra of functions on the phase space 
is {\em not\/} a common structure of classical and quantum mechanics, 
being given by {\em a quotient} of the common Poisson algebra $\LRM$.

\bsk\noindent
{\em The non-commutative Poisson algebra generated by cartesian coordinates and
momenta}.

\msk\noindent
As discussed above, the r\^ole of the Lie-Rinehart relations is mainly 
that of allowing for the identification of $\LRM$ with the algebra of 
mechanical variables on a manifold. If only $\R^n$, with cartesian coordinates, 
is considered, a similar simplified construction 
still yields a Poisson algebra with a central element $Z$ satisfying 
eqs.(\ref{Z1}), (\ref{Z2}).

For its construction, it is enough to consider the polynomial algebra in the
cartesian coordinates $x_i , i = 1 \ldots n$ and the Lie algebra $\L_c$
generated by it and by momenta $p_i , i = 1 \ldots n$ with Lie product
\be
\{ \, P(x) , \, p_i\}  = \frac{\partial}{\partial x_i } P(x) \ , \ \ \
\{ \, p_i , \, p_j \} = \{ \, P_1(x) , \, P_2(x) \} = 0 \ .
\ee
No Rinehart product  between coordinates and momenta is present, since only
the vector fields associated to translations are considered.
Then, Definition 1, without condition iv), gives a unique 
universal enveloping Poisson algebra $\Lambda_c$
of $\L_c$, extending the (commutative) algebraic structure of 
polynomials in the coordinates.
The proof of Theorem 1 shows that
$ \, [\, x_i, \, p_i\, ] \,  $ is independent of $i$ 
and defines an element $Z \in \Lambda_c$ satisfying 
eqs.(\ref{Z1}), (\ref{Z2}).

A non-trivial point is that $\Lambda_c$ is not an
explicit polynomial algebra; rather, it is uniquely determined
by the requirements that 
i) it is a Poisson algebra, 
ii) it is generated by the polynomials in $x_i$ and by the momenta $p_i$,
iii) it is universal, i.e. for any Poisson algebra $\Lambda$ 
satisfying i) and ii) there is a unique homomorphism 
$\rho: \Lambda_c \mapsto \Lambda$ acting as the identity on the generators. 

The same Poisson algebra is also obtained if one only starts with 
the {\em Lie} algebra $\L_H$ generated by $x_i$, $p_i$ 
and an element $I$, satisfying
\be 
\{ \, x_i , \, p_j \} = \delta_{ij} \, I \, , \ 
\ee
all the other Lie products vanishing. One considers the
universal enveloping Poisson algebra given by Definition 1, dropping 
conditions iii) and iv) and keeping ii) for $I$, i.e. with $I$ as the 
identity. Then, introducing as before $Z \equiv [x_i, p_i]$ ($i$ fixed)
abelianess of the polynomials in $x_i$ follows from eq.(\ref{Z1})
and the result is again the Poisson algebra $\Lambda_c$.

Taking in $\Lambda_c$ the quotients defined by the ideals
generated by $Z^* Z - z^2$, $z \geq 0$ one obtains  
the classical Poisson algebra of polynomials in coordinates and momenta
and the Heisenberg algebra, with $\hbar = z$. 

The Dirac ansatz for commutators between cartesian coordinates 
and momenta has therefore no alternative, precisely in the sense that 
the Heisenberg algebra and the classical polynomial Poisson algebra are
the only Poisson algebras which envelope 
$\L_c$ or $\L_H$ in the above sense 
(and are therefore isomorphic to quotients in $\Lambda_c$) 
and represent  $Z^*Z$ by a nonnegative number.

A more complete discussion of central variables and ideals 
in Poisson algebras requires the introduction of bounded 
variables, which can be conveniently constructed 
in Hilbert space representations.

\bsk\noindent
{\em Regular factorial representations of $\LRM$. 
Classical and Quantum Mechanics}.

\msk\noindent
\begin{definition}{Definition} 
A {\bf representation} $\pi$ of a Poisson *-algebra $\Lambda$ in a complex 
Hilbert space $\H$ is a  homomorphism of $\Lambda$ into a Poisson 
*-algebra of operators in $\H$, with both the operator product 
and a Lie product $\{.\,, \,.\}$ satisfying the
Leibniz rule, having a common invariant dense domain $D$ on which  
$$\pi(A
\cdot B) = \pi(A) \,\pi(B), \,\,\,\,\pi(\{\,A,\,B\,\}) =
\{\,\pi(A),\,\pi(B)\,\}, \,\,\,\,\pi(A^*) = \pi(A)^*.$$

\noindent
A representation $\pi$ of $\LRM$  is called 
{\bf regular} if $\pi(C_0^\infty(\M)) \neq 0$ and 
\newline i) ({\bf exponentiability}) 
$D$ is invariant under  
$\pi(\cifm)$ and the one parameter unitary groups $\Ulv$, $\Ulz $,
$\lambda \in \R$, generated by $T_v \equiv \pi(v)$ and $ - i\, \pi(Z)$, 
respectively,
\newline ii) ({\bf diffeomorphism invariance}) the elements $\glv \in \G(\M)$
define strongly continuous automorphisms of the $C^*$ algebra $\gotA_\pi$  
generated by $\pi(\cifm)$, $\Ulv$, $\Ulz $,
\be
\gmw: \pi(f) \ra \pi(\gmw f),\,\,\,\, 
\Ulv \ra U(\lambda \gmw(v)), \,\,\,\,\, \Ulz \ra \Ulz.
\ee

\noindent
A regular representation $\pi$ of $\LRM$ is called {\bf factorial} if the elements
of  the center ${\cal Z}_\pi$ of $\gotA_\pi'' $, the weak closure of $\gotA_\pi$, 
which are invariant under Diff($\M$) are multiples of the identity.
\end{definition}

Condition i) requires the existence of exponentials of the
representatives of the vector fields and of $Z$; such
exponentials are unique since stability of $D$ under them implies 
essential selfadjointness of the generators on $D$.
Condition ii) amounts to exponentiability of the derivations 
defined by eq.(\ref{ActDiff}), in the representation $\pi$; 
it is implied by i) for representations with $z \neq 0$ 
(as a consequence of eq.(\ref{CP}) below).
The action of diffeomorphisms on $\gotA_\pi'' $ is well defined as a
consequence of their strong continuity (condition ii). 

The above condition on the center of $\gotA$ reflects the fact that 
$\LRM$ has both an associative product and a Lie product, related to
Diff($\M$) by eq.(\ref{ActDiff}), so that diffeomorphism invariance 
of an element corresponds, in exponentiated form, to the vanishing of its Lie
brackets with vector fields. For representations with $z \neq 0$,
central elements are automatically diffeomorphism invariant, 
by eq.(\ref{CP}) below.

For the analysis of regular factorial representations of $\LRM$,
one has \cite{MS2}:

\begin{lemma}{Proposition} In a representation $\pi$ of $\LRM$,
$\pi(f)$ and $\pi(v)$, $f \in \cifm$, $v \in$ Vect($\M$),
are strongly continuous on $D$ in the $C^\infty$ topology of $\cifm$ and
Vect($\M$). Equation (\ref{ActDiff}) holds for $\pi(\glv(A))$, $A = f,\, v$, 
with the derivative taken in the strong topology.

\noindent
In a regular representation, the one-parameter unitary groups 
$\Ulv$, $\Ulz $ satisfy 
\be
 [\,\Ulv, \,\Ulz\,] = 0, \,\,\,\,\,\,[\,\pi(f), \,U(\lambda Z)\,] =0.
\ee

\noindent
In a regular factorial representation one has 
\newline i) $\Ulz = e^{- i \lambda z} I $, $z \in \R$; modulo the $ ^*$
involution in $\LRM$  (leaving $C^\infty(\M) +$ Vect ($\M$) 
pointwise invariant), one can take $ z \geq 0$,
\newline ii) the one parameter groups $\Ulv$ are strongly continuous in $v$ in
the $C^\infty$ topology of the vector fields and
\be
U(\lambda v)\,U(\mu\,w) = U(\mu g_{\lambda z v}(w)) \,U(\lambda v), \,\,\,\,\,\,\,\,
 U(\lambda v) f  = \glzv(f)\, U(\lambda v) \, , 
\label{CP}
\ee  
\end{lemma}

For $z \neq 0$, eq.(\ref{CP}) defines, with the obvious modification
of a factor $z$ in the Lie algebra structure constants, the crossed 
product $\Pi(\M)$ of $\cifm$ and $\tilde \G(\M)$, 
the universal covering group of Diff($\M$)
(the usual definition corresponding to $z=1$). 
A regular representation of $\LRM$ gives a representation of $\Pi(\M)$ which
is Lie-Rinehart regular in the sense of \cite{MS2}, since it is differentiable,
the generators are strongly continuous in the $C^\infty$ topology of 
vector fields and they satisfy the Lie-Rinehart relations.

We recall that two representations are called quasi equivalent if each of 
them is unitarily equivalent to a sum of subrepresentations of the other.
Our main result is that the regular factorial representations of $\LRM$ 
exactly define classical and quantum mechanics on $\M$, with $z$ playing 
the r\^ole of $\hbar$:

\begin{theorem}{Theorem} 
\cite{MS2}  The regular factorial representations  $\pi$ of $\LRM$ are
classified, modulo the $ ^*$  involution, 
by the values \, $i z, \, z\geq 0$ of the central variable 
$Z$ and  \vspace{0.5mm}
\newline 1) for $z > 0$, they coincide, apart from a
multiplicity, with of the irreducible Lie-Rinehart regular 
representations of the crossed product 
$\Pi(\M) \equiv \cifm \times \tilde \G (\M)$, defining \ {\bf Quantum Mechanics}
on $\M$. As a result of \cite{MS1}, for each $z>0$, they are 
locally equivalent, up to a multiplicity, to the Schroedinger 
representation and they are classified by the unitary representations of 
the fundamental group of $\M$. 
\vspace{0.5mm}
\newline 2) for z = 0, for  separable representation spaces $\H$, they are
quasi equivalent to the representation $\pi_C$ in
$L^2(T^* \M, d x \,d p)$, defined by multiplication operators 
({\bf Classical Mechanics}): on 
$D = C_0^\infty(T^* \M)$, in local coordinates, 
$\forall f \in \cifm$, $\forall v = \sum_i g_i(x) \partial/\partial x_i$, 
supp\,$v \subset \O$, $\O$ homeomorphic to  an open  disc,  
\be
 \pi_(f) = f(x), \ \ \ \
 \pi(v) = \sum_i g_i(x)\, p^i  \ ,
\ee
$p^i$ denoting the coordinates in the basis dual to $\partial/\partial x_i$.
The Lie product in $\pi_C(\LRM)$ is given by the standard
Poisson brackets on $T^* \M$.
If $\difm$ is unitarily implemented, the representation
is unitarily equivalent to a multiple of $\pi_C$.
\end{theorem}

\noindent
{\em Proof}. 
By i) of Proposition 1, $Z = iz I$ in regular factorial representations of
$\LRM$ and for $z \neq 0$ the classification follows from Proposition 1 and 
Theorems 3.7, 4.5, 4.6 of ref.\cite{MS1}. 

For $z = 0$, by separability of $\H$, modulo unitary equivalence, 
the representation is defined by multiplication operators 
in a denumerable sum of $L^2$ spaces over the spectrum of the abelian 
$C^*$ algebra $\gotA$. 

The proof \cite{MS2} then  requires three steps: first, the spectrum of 
$\gotA$ is identified, apart from a set of zero measure, with the 
cotangent bundle $T^*(\M)$; in fact, by regularity of $\pi$, 
almost all the multiplicative functionals 
$\xi$ on $\gotA$ are determined by their value on $\cifm$ and 
on the generators $\pi(v)$ of the one parameter groups, to 
which they extend by regularity; locally, 
\be
\xi (v) \equiv \xi \, (\sum_i g_i(x) \frac {\partial}{\partial x_i}) = 
  \sum_i g_i(x_\xi) \, \xi (\frac {\partial}{\partial x_i}) \equiv
  \sum_i g_i(x_\xi) \, p^i_\xi \ ,
\ee 
since $\M$ is the spectrum of the closure of $\cifm$; therefore, 
$\xi = (x_\xi, p^i_\xi) \in T^*(\M)$ and
\be
\pi(f)= f(x_\xi) \ , \ \ \ \pi(v) =  \sum_i g_i(x_\xi) \, p^i_\xi 
\ee
as multiplication operators.

The second point is the regularity of the above measures with respect to the
Lebesgue measure on $T^*(\M)$, which follows using transitivity of the 
transformations of $T^*(\M)$ induced by diffeomorphisms of $\M$ (apart from
a set of zero measure) and local coordinates defined (almost everywhere) 
in $T^*(\M)$ by suitable vector fields on $\M$.  

Third, the identification of the Lie brackets with the classical
Poisson brackets on $\LRM$ follows, by the Leibniz rule, from its validity
for $\pi (\cifm + $  Vect($\M$)); from Proposition 1,
one has, in local coordinates, $\forall A(x, p) \in \pi(\cifm + $
Vect($\M$)), 
$v = \sum_i g_i(x) \, p^i$,  $\, \glv(x,p)$ the canonical transformation defined
on $T^*(\M)$ by the diffeomorphism $\glv$, 

\be
\{\sum_i g_i(x) \,p^i,\,A(x, p)\,\} = 
(d/ d \lambda) A(\glv^{-1}(x, p))|_{\lambda = 0} =
\ee
\be
  = \sum_i \Big{(} - \frac{\partial A(x, p)}{\partial x_i}\, g_i(x) +
\frac{\partial A(x, p)}{\partial p^i}\,\frac{\partial g_j(x)}{\partial x_i}\,
p^j \Big{)} = \{\sum_i g_i(x) \,p^i,\,A(x, p)\,\}_{Class} \ .  
\ \ \rule{5pt}{5pt} 
\ee

\end{document}